\documentclass[12pt]{article}
\usepackage{amsmath}
\usepackage{amsfonts}
\usepackage{graphicx}
\newcommand{\bean}{\begin{eqnarray*}}
\newcommand{\eean}{\end{eqnarray*}}
\newcommand{\ed}{\end{document}}

\newcommand{\be}{\begin{equation}}
\newcommand{\ee}{\end{equation}}
\newcommand{\barr}{\begin{array}}
\newcommand{\earr}{\end{array}}
\newcommand{\bea}{\begin{eqnarray}}
\newcommand{\eea}{\end{eqnarray}}

\begin{document}
\title{Solving the Einstein field equations}
\author{A.V.Bratchikov
\\ Kuban
State Technological University,\\ 2 Moskovskaya Street, Krasnodar,
  350072, Russia
\\
}
\date{} \maketitle
\begin{abstract}
A solution of the Einstein vacuum field equations is constructed within the contex of perturbation theory. The solution possesses a graphical representation in terms of diagrams. 
\end{abstract}



In a recent article \cite {B1} for a wide class of nonlinear equations a perturbative solution was constructed.This class includes equations of motion of field theories. Solutions of the pure Yang-Mills field equations and equations of motions in spinor electrodynamics \cite{B2} were obtained. The solutions can be described by diagrams.
The aim of the present paper is to 
construct a solution of the Einstein field equations in the vacuum.

Let $x=(x^0,x^1,x^2,x^3)
$ be space-time coordinates, and $g(x)=g_{\mu\nu}(x)dx^\mu dx^\nu$ a space-time metric. The vacuum Einstein equations read 
\bea \label{Gr}
G_{\mu\nu}= 0,
\eea
where 
\bean
G_{\mu\nu}= R_{\mu\nu}-\frac 1 2 g_{\mu\nu}R,
\eean
$R_{\mu\nu}$ is the Ricci tensor
\bean \label{A}
R_{\mu\nu}=\Gamma^\beta_{\mu\nu,\beta}-\Gamma^\beta_{\mu\beta,\nu}+\Gamma^\beta_{\mu\nu}\Gamma^\alpha_{\beta\alpha}-\Gamma^\alpha_{\mu\beta}\Gamma^\beta_{\nu\alpha},
\eean
$R=R^\mu_\mu,$
and $\Gamma_{\alpha\beta}^\gamma$ is the Cristofel symbol 
\bean \label{A}
\Gamma_{\alpha\beta}^\gamma=\frac 1 2 g^{\gamma\rho}(\partial_\beta g_{\rho\alpha}+\partial_\alpha g_{\rho\beta}-\partial_\rho g_{\alpha\beta}).
\eean

For $\eta=\eta_{\mu\nu}dx^\mu dx^\nu,\eta_{\mu\nu}=diag (1,-1,-1,-1),$  we can expand equations (\ref{Gr}) in powers of $h=g-\eta$
\bea  \label{E}
\sum_{n=1}^\infty 
G^n_{\mu\nu}(h)=0.
\eea
Here 
\bean  \label{b}
G^n_{\mu\nu}(h)=\left.\frac 1 {n!}
\frac {d^n G_{\mu\nu}(\eta+\xi h)} 
{d^n\xi}\right|
_{\xi=0}
.
\eean
We impose a gauge condition 
\bea \label{b}\eta^{\alpha\beta}\partial_\alpha \varphi_{\beta\nu}=0,
\eea 
where 
\bea  \label{bl}\varphi_{\mu\nu}= h_{\mu\nu}-\frac 1 2\eta_{\mu\nu}h_\alpha^\alpha, \qquad h_\alpha^\alpha=\eta^{\alpha\beta}h_{\alpha\beta}.
\eea 
From equation (\ref{bl}) it follows
\bea\label{le}h_{\mu\nu}=\varphi_{\mu\nu}-\frac 1 2 \eta_{\mu\nu}\varphi^\alpha_\alpha,\qquad\varphi^\alpha_\alpha=\eta^{\alpha\beta}\varphi_{\alpha\beta}.
\eea

Let $\tilde G^n(\varphi)$ be defined by $\tilde G^n(\varphi)= G^n(h)_{\mu\nu}dx^\mu dx^\nu,$
$\varphi =\varphi_{\mu\nu}dx^\mu dx^\nu.$  
Then equation (\ref {E}) takes the form \cite{LL} 
\bea  \label{A}
\Box \varphi-2\sum_{n=2}^\infty 
\tilde G^n(\varphi)=0,
\eea
where $\Box=\eta^{\alpha\beta}\partial_\alpha \partial _\beta.$ 

The general solution $
\varphi_{0\mu\nu} 
$ of the free equation
\begin{eqnarray*} \label{U}
\Box \varphi_{\mu\nu}=0
\end{eqnarray*}
is given by \cite {CG}
\begin{eqnarray*}
\varphi_{0\mu\nu}= \frac 1 {4\pi}M(v_{\mu\nu})+\frac 1 {4\pi}\frac \partial {\partial t}(tM(u_{\mu\nu})),
\end{eqnarray*}
where 
\begin{eqnarray*}
M(w_{\mu\nu})= \int_{S} 
w_{\mu\nu}( x^1+t\xi^1, x^2+t\xi^2, x^3+t\xi^3) d \sigma_\xi, 
\end{eqnarray*}
$\xi^1, \xi^2,\xi^3$ are coordinates on the unit sphere $S$, $\sigma_\xi$ is the area element on $S,$  $t=x^0,$
\begin{eqnarray*}\varphi_{0\mu\nu}(0,x^1,x^2,x^3)=u_{\mu\nu}(x^1,x^2,x^3),
\end{eqnarray*}
\begin{eqnarray*}
 \left.\frac {\partial \varphi_{0\mu\nu}(t,x^1, x^2,x^3)}{\partial t}\right|_{t=0}=v_{\mu\nu}(x^1, x^2,x^3),
 \end{eqnarray*}
and $u_{\mu\nu},v_{\mu\nu}$ are some functions.
Equation (\ref {b}) implies 
\begin{eqnarray*}v_{0\nu}-\sum_{i=1}^3 \partial_iu_{i\nu}=0.
\end{eqnarray*}

A specific solution  ${\cal E}(f_{\mu\nu})$ of the inhomogeneous equation
\begin{eqnarray*} \label{U}
\Box \varphi_{\mu\nu}=f_{\mu\nu}
\end{eqnarray*}
reads  
\begin{eqnarray*}
{\cal E}(f_{\mu\nu})=\frac 1 {4\pi}\int^t_0\tau d\tau \int_{S} 
f_{\mu\nu}(t-\tau, x^1+\tau \xi^1, x^2+\tau \xi^2, x^3+\tau \xi^3) d \sigma_\xi. 
\end{eqnarray*} 

Then equation  (\ref {A}) can be written in the form 
\bea \label{bas}
\varphi =\kappa \varphi_0+\sum_{n=2}^\infty p_n(\varphi),
\eea
where $
\varphi_0=
\varphi_{0\mu\nu}dx^\mu 
dx^\nu,$ and $p_n
=2{\cal E} ( 
\tilde G^n_{\mu\nu}) dx^\mu dx^\nu.$ Here we introduced an auxiliary parameter $\kappa$. Eventually it is set to be $1.$

Let $V$ be the space of symmetric 2-forms. 
We shall need the polylinear symmetric functions $$\langle \ldots  \rangle_n:
{V }^n 
 \to V,\quad n=2,3,\ldots,$$
defined for $\gamma_1,\ldots,\gamma_n\in V$ by 
\begin{eqnarray} \label {orsk}
\langle \gamma_1,\ldots,\gamma_n \rangle_n = \frac \partial {\partial a_1}\ldots \frac \partial {\partial a_n}p_n( a_1\gamma_{1}+\ldots+ a_n \gamma_{n}).
\end{eqnarray} 
One can check that
\bean \label {or}
\langle \gamma,\ldots,\gamma \rangle_n = {n!} p_n(\gamma). 
\eean 
Then equation (\ref{bas}) becomes 
\bea \label {equv}
\varphi= \kappa\varphi_0+\sum_{n=2}^\infty \frac 1 {n!} \langle \varphi,\ldots,\varphi \rangle_n.
\eea

For $m\geq 2,1\leq i_1< \ldots < i_n\leq m$ let $$P^m_{i_1\ldots i_n}: V^m\to V^{m-n+1} $$ be defined by
\bean \label {or}
P^m_{i_1\ldots i_n}( \gamma_1,\ldots ,\gamma_m ) = 
( \langle \gamma_{i_1},\ldots, \gamma_{i_n}\rangle_n ,\gamma_1,\ldots,\widehat{\gamma}_{i_1},\ldots,\widehat{\gamma}_{i_n},\ldots,\gamma_{m} ),
\eean 
where $\widehat{\gamma}$ means that $\gamma$ is omitted.
If $\phi\in V$ is given by  
\bea \label {v} \phi = P^{n_s}_{I_s}\ldots P^{m-n_1+1}_
{I_2}
P^{m}_{I_1}
(\gamma_1,\ldots ,\gamma_m )
\eea
for some $I_1= (i_1^1\ldots  i^1_{n_1}),I_2=(i^2_1\ldots i^2_{n_2}),\ldots,I_s=(i^s_1\ldots i^s_{n_s}),
$
we say that  $\phi$ is a descendant of $(\gamma_1,\ldots ,\gamma_m ).$

Each  descendant can be represented by a diagram. 
In this diagram an element of  $V$ is represented by the line segment \rule[3pt]{20pt}{0.5pt}\,\,. A product $$(\gamma_1,\ldots,\gamma_{n})\to\langle \gamma_1,\ldots,\gamma_{n}\rangle_n$$ is represented by the vertex joining the line segments for  $ \gamma_1,\ldots,\gamma_{n},$ $ \langle \gamma_1,\ldots,\gamma_n\rangle_n.$ 
The general rule for graphical representation of $P^{m}_{I}
(\gamma_1,\ldots ,\gamma_m )$ should be clear
from Figure 1.Here we show the diagram for $$P^m_{ijk}(\gamma_1,\ldots,\gamma_m)=( \langle \gamma_{i},\gamma_j,\gamma_k\rangle_3 ,\gamma_1,\ldots,\widehat{\gamma}_{i},\ldots,\widehat{\gamma}_{j},\ldots,\widehat{\gamma}_{k},\ldots,\gamma_{m} ).$$
The points labeled  by $1,\ldots, m$ represent the ends of the lines for $\gamma_1,\ldots ,\gamma_m.$  Using the representation for $P^m_{i_1\ldots i_n}( \gamma_1,\ldots ,\gamma_m )$ one can consecutively draw the diagrams for 
$P^{n_1}_{I_1}(\gamma_1,\ldots ,\gamma_m ),P^{n_2}_{I_2}
P^{n_1}_{I_1}(\gamma_1,\ldots ,\gamma_m ),\ldots , \phi$ (\ref {v}). 
\begin{figure}\centering\includegraphics[width=1.75in]{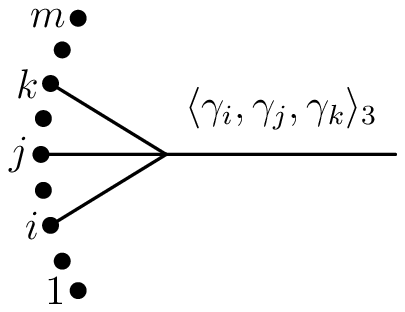}\begin{center} Figure 1. Diagram for $P^m_{ijk}(\gamma_1,\ldots,\gamma_m).$\end{center}\end{figure}

Let us introduce a family of functions $$\langle \ldots  \rangle:
{V }^m 
 \to V,\qquad m=2,3,\ldots,$$
such that for $\gamma_1,\ldots,\gamma_m \in V$ $\langle \gamma_1,\ldots,\gamma_m \rangle$ is 
defined as the sum of all the descendants of its arguments.
For example, for $m=2$ and $m=3$ we have
\bean \label {or}
\langle \gamma_1,\gamma_2 \rangle = \langle \gamma_{1},\gamma_{2}\rangle_2,
\eean 
\bean \label {or}
\langle \gamma_1,\gamma_2,\gamma_{3}\rangle=
\langle \langle \gamma_{1},\gamma_{2}\rangle_2, \gamma_3 \rangle_{2}+\langle \langle \gamma_{1},\gamma_{3}\rangle_2, \gamma_2 \rangle_{2}+\langle  \langle \gamma_2, \gamma_3 \rangle_2,\gamma_{1} \rangle_2+\langle \gamma_1, \gamma_2,\gamma_3 \rangle_3. 
\eean

A solution of equation (\ref {equv}) is given by \cite{B1}
\bean \label {orro}
\varphi= \langle e^{\kappa \varphi_0} \rangle, 
\eean
where
\bean \label {}
\langle e^{\kappa \varphi_0} \rangle = \sum_{n=0}^\infty \frac {\kappa^n} {n!}\langle 
\varphi_0^n \rangle
,
\eean
$\langle 
\varphi_0^n\rangle$ is defined as $0$ if $n=0,$ and otherwise $$\langle 
\varphi_0^n\rangle=\langle 
\underbrace{\varphi_0,\ldots ,\varphi_0}_{\text {n times}}\rangle,\qquad \langle \varphi_0\rangle=\varphi_0.$$

For example, the $O(\kappa^3)$ contribution in $\varphi$ is given by  
\begin{eqnarray*}
\frac {\kappa^3} 6\langle 
\varphi_0,
\varphi_0, 
\varphi_0\rangle
=\kappa^3\left (\frac {1} 2
\langle\langle \varphi_0, \varphi_0\rangle_2, \varphi_0\rangle_2+\frac 1 6\langle 
\varphi_0,
\varphi_0, 
\varphi_0\rangle_3\right).
 \end{eqnarray*}
The diagram for $ \langle\langle \varphi_0, \varphi_0\rangle_2, \varphi_0\rangle_2 $ is depicted in 
Figure 2. 

\begin{figure}
\centering
\includegraphics[width=3in]{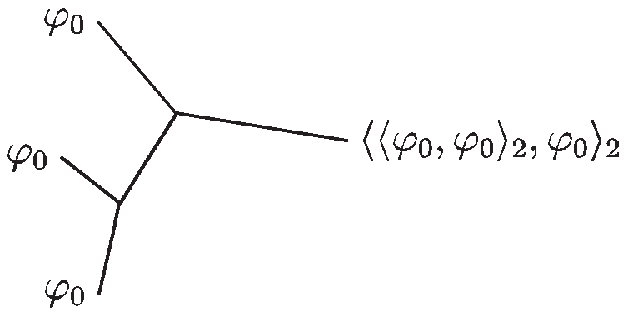}
\begin{center}
Figure 2. Diagram for $\langle\langle \varphi_0, \varphi_0\rangle_2, \varphi_0\rangle_2.$
\end{center}
\end{figure}
\newpage

The tensor $h_{\mu\nu}$ can be found by using (\ref {le}).

\end{document}